\documentclass[aps,prl,twocolumn,superscriptaddress,showpacs,preprintnumbers]{revtex4}
\usepackage{graphicx}
\usepackage{amssymb}

\newcommand{\slrr}      {$T_1^{-1}$}

\newcommand{\lescox}    {La$_{1.8-x}$Eu$_{0.2}$Sr$_{x}$CuO$_4$}
\newcommand{\ybcox}     {YBa$_{2}$Cu$_{3}$O$_{7-\delta}$}

\newcommand{\lrescox}    {La$_{2-x-y}$RE$_{y}$Sr$_{x}$CuO$_4$}
\newcommand{\lscox}     {La$_{2-x}$Sr$_{x}$CuO$_4$}

\newcommand{\oxy}       {$^{17}$O}

\begin{document}


\preprint{LA-UR-05-1085}

\title{NMR Evidence for Charge Inhomogeneity in Stripe Ordered \lescox}
\author{H.-J. Grafe}
\affiliation{Condensed Matter and Thermal Physics, Los Alamos
National Laboratory, Los Alamos, NM 87545, USA}
\affiliation{Leibniz-Institut f\"{u}r Festk\"{o}rper- und
Werkstoffforschung, Dresden, Helmholtzstr. 20, 01171 Dresden,
Germany}
\author{N. J. Curro}
\affiliation{Condensed Matter and Thermal Physics, Los Alamos
National Laboratory, Los Alamos, NM 87545, USA}
\author{M. H\"{u}cker}
\affiliation{Physics Department, Brookhaven National Laboratory,
Upton NY, USA}
\author{B. B\"{u}chner}
\affiliation{Leibniz-Institut f\"{u}r Festk\"{o}rper- und
Werkstoffforschung, Dresden, Helmholtzstr. 20, 01171 Dresden,
Germany}

\date{\today}

\begin{abstract}
We report \oxy\ Nuclear Magnetic Resonance (NMR) results in the
stripe ordered \lescox\ system. Below a temperature $T_q \sim
80$K, the local electric field gradient (EFG) and the absolute
intensity of the NMR signal of the planar O site exhibit a
dramatic decrease.  We interpret these results as microscopic
evidence for a spatially inhomogeneous charge distribution, where
the NMR signal from O sites in the domain walls of the spin
density modulation are wiped out due to large hyperfine fields,
and the remaining signal arises from the intervening Mott
insulating regions.
\end{abstract}



\maketitle

Several doped transition metal oxides exhibit inhomogeneous charge
stripe order on a mesoscopic scale due to competing long and short
range interactions acting on the charge carriers
\cite{gorkov,bishop}. In the cuprates, the doped charge carriers
(holes) are expected to form one-dimensional channels (charge
stripes) separating regions of insulating antiferromagnetic order
of the Cu spins (spin stripes) \cite{zaanen}. The presence of this
inhomogeneity may be vital to the mechanism of d-wave
superconductivity \cite{kivelson}, however direct experimental
evidence for such structures has been elusive. To date, the only
observations have been via techniques that probe either the spin
density or the charge inhomogeneity: Neutron Scattering (NS)
experiments provided the first evidence for the modulation of spin
density that is expected in a stripe lattice
\cite{tranquadaNature}, whereas NMR and Scanning Tunnelling
Microscopy (STM) experiments indicate the presence of
inhomogeneous doping distributions
\cite{haaseslichter,haase2,singer,pan,yazdani}. In this Letter we
discuss new NMR data that provide direct evidence for a
correlation between the local charge and spin density maps by
taking advantage of the unique properties of the planar oxygen to
probe simultaneously both the local spin structure as well as the
local hole doping in the O p-orbitals. Our data reveal that not
only is the charge spatially inhomogeneous, but that the regions
of excess charge are correlated with the domain walls of the spin
order, exactly as expected for a stripe pattern
\cite{zaanen,scalapino}.

The rare-earth co-doped \lescox\ series is ideal for NMR
investigations of the spin and charge inhomogeneity. Structurally,
this material is almost identical to the prototypical high
temperature superconductor \lscox, but undergoes a subtle phase
transition to the low temperature tetragonal (LTT) structure below
$T_{\rm LT}$ = 135K \cite{klauss}.  Instead of superconducting
below $T_c\sim$ 35K, this system exhibits glassy magnetic order
below $T_N\sim$ 25K
\cite{klauss,kataev,curroPRL,kataev2,teitelbaumwipeout,
teitelbaum3sites,simovic}, and elastic NS measurements have
identified static long-range spin and structural modulations that
are likely produced by stripe order \cite{tranquadaNature}.  The
slow spin fluctuations in this system dominate the NMR response of
the La and Cu nuclei \cite{curroPRL,simovic,hunt}.  However, the
planar oxygen does not suffer the same fate: it experiences an
isotropic transferred hyperfine coupling (129 kOe/$\mu_{\rm B}$)
to the two nearest neighbor Cu spins, so for antiferromagnetically
correlated neighbors, the hyperfine field at the O site vanishes
\cite{ishida}. Furthermore,  \oxy\ ($I$=5/2) has a quadrupolar
moment ($^{17}Q$ = -2.56 $\times 10^{-26}$cm$^2$), so it is
sensitive to the EFG at the nuclear site, which is a measure of
the local hole doping \cite{haase}. Although previous NMR studies
have shown the presence of spatial modulations due to doping
inhomogeneities in superconducting \lscox\
\cite{haaseslichter,singer}, O NMR in \lescox\ provides a unique
opportunity to investigate the doping inhomogeneity in a system
where the spin fluctuations are suppressed.

\begin{figure}
\includegraphics[width=38mm]{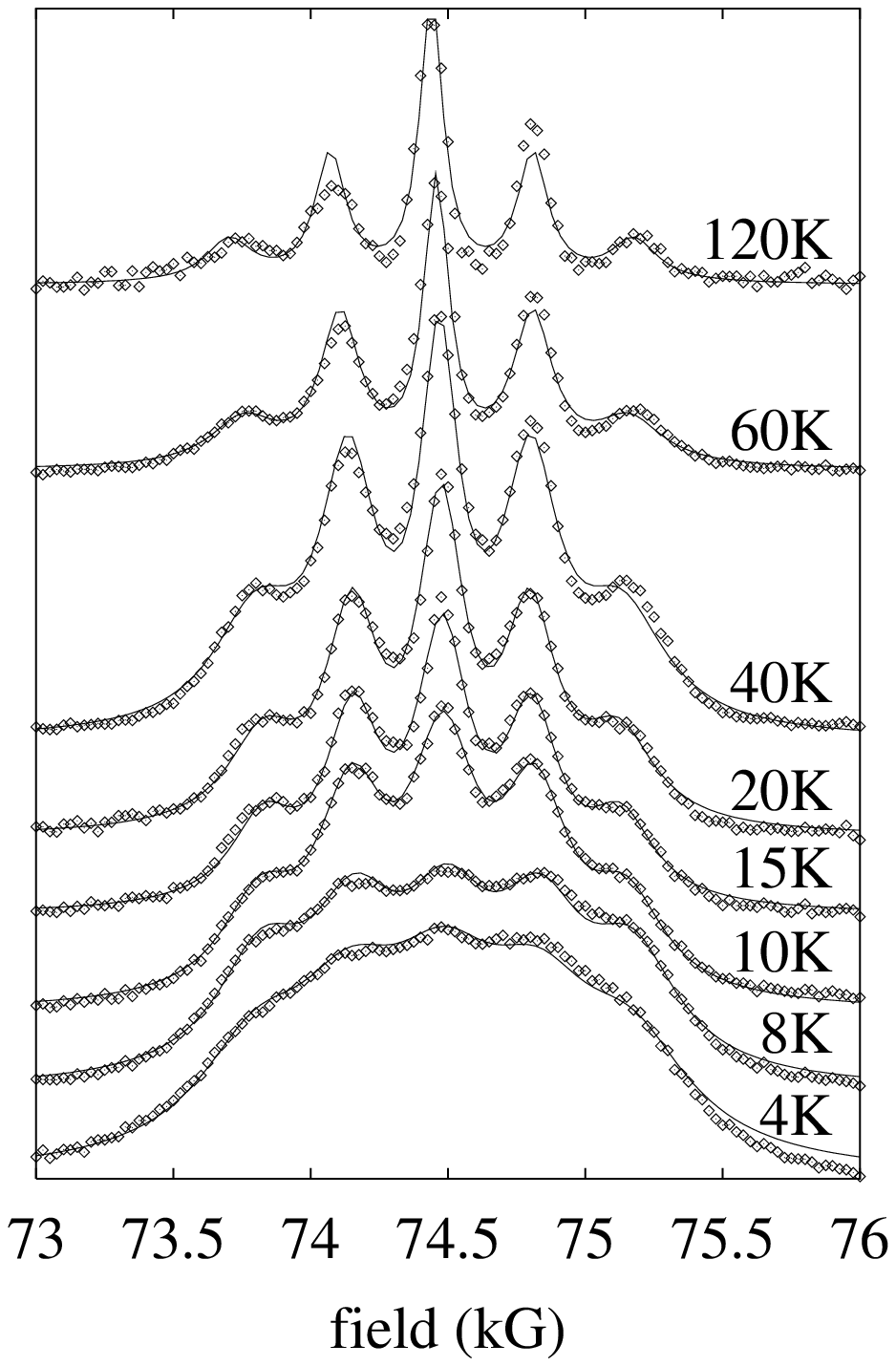}
\includegraphics[width=38mm]{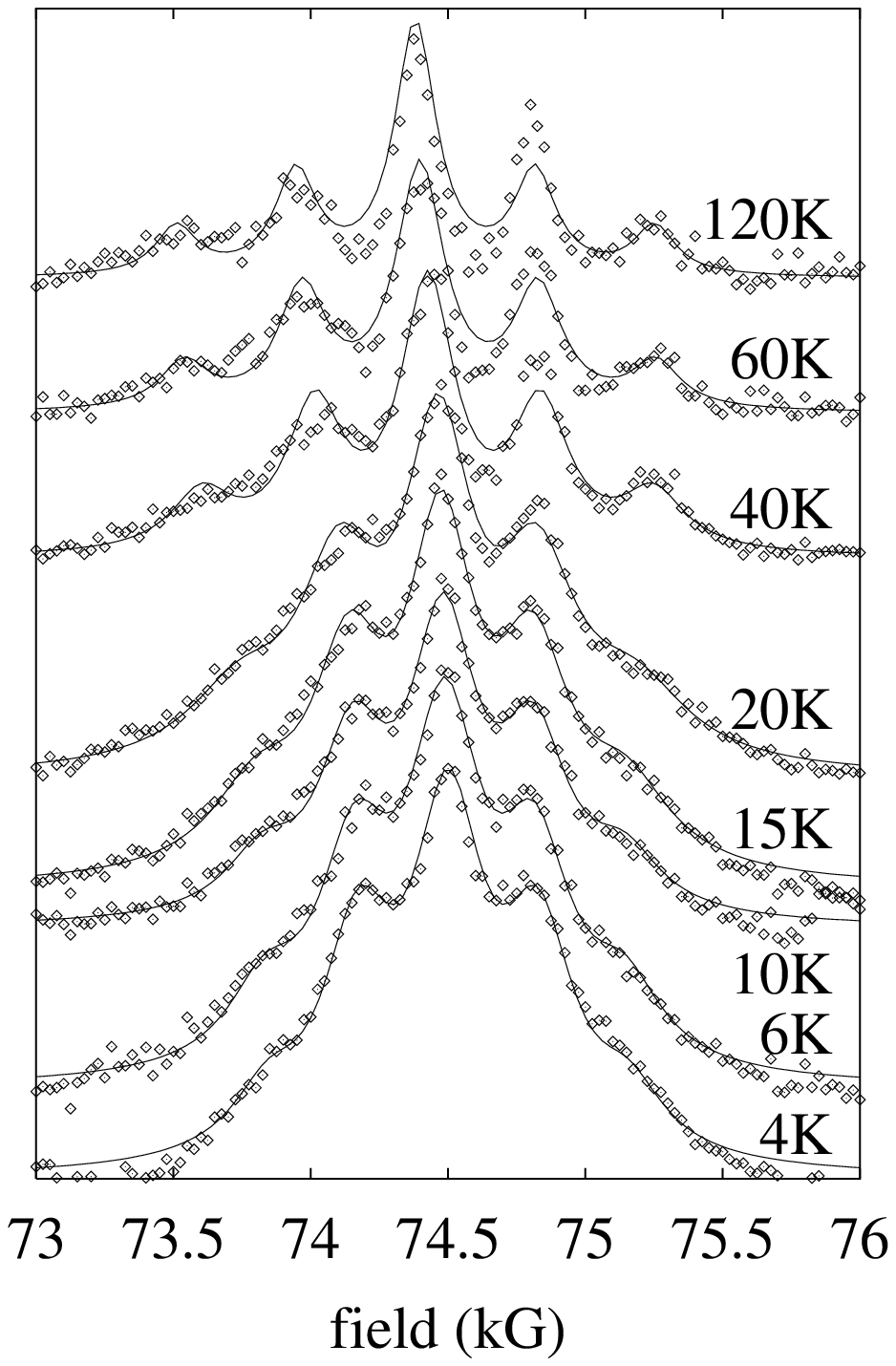}
\caption{\label{fig:spectra} NMR field-swept spectra of the planar
oxygen in \lescox\ for $x=0.13$ (left) and $x=0.2$ (right) at 43
MHz. The solid lines are fits as described in the text. All
spectra are normalized to equal heights for comparison.}
\end{figure}

Ground polycrystals of \lescox\ with $x$=0.08, 0.105, 0.13, 0.17
and 0.2 were enriched with \oxy\ by annealing in \oxy$_2$ gas at
600$^{\circ}$C for 24 hours. The powder samples were then mixed
with epoxy and aligned along the $c$-axis by curing in an external
field. The \oxy\ NMR spectra were obtained by measuring the spin
echo intensity while sweeping the magnetic field along the
$c$-axis at fixed frequency. $^{139}$La spectra were independently
measured in non-enriched samples and were subtracted from the
spectra of the enriched samples to obtain the data in Fig
(\ref{fig:spectra}). The spectra clearly show five transitions
split by the quadrupolar interaction with a value consistent with
that of the planar O in \lscox\ \cite{apicalnote}. The Hamiltonian
is given by:
\begin{equation}
\mathcal{H}=\gamma\hbar\hat{I}\cdot\mathbf{H}_0+\frac{h\nu_c}{6}(3\hat{I}_z^2-I^2+\eta(\hat{I}_x^2-\hat{I}_y^2))+\mathcal{H}_{\rm
hyp}
\end{equation}
where $\gamma$ is the gyromagnetic ratio, $\mathbf{H}_0$ is the
external field, $\nu_{c} = 3eQV_{cc}/20$,
$\eta=(V_{aa}-V_{bb})/V_{cc}$, $Q$ is the quadrupolar moment of
the \oxy\ and $V_{\alpha\alpha}$ are the components of the EFG
tensor. The hyperfine interaction is give by $\mathcal{H}_{\rm
hyp} = C\hat{I}\cdot\sum_{i\in nn}\mathbf{S}(\mathbf{r}_i)$, where
$C=129$kG/$\mu_B$ and the sum is over the two nearest neighbor Cu
spins \cite{ishida}.  In the absence of static magnetic order, the
resonance field of each transition at fixed frequency $f$ is given
by $\gamma H_n = (f-n\cdot\nu_c)/(1+K_c)$, where $n$=-2,-1,0,1 or
2, and $K_c$ is the Knight shift. The spectra were fit to
Lorentzian distributions centered at the $H_n$ with widths $\sigma
= \sqrt{\sigma_m^2+(n\cdot\sigma_q)^2}$, where $\sigma_m$ is the
magnetic linewidth and $\sigma_q$ the quadrupolar linewidth caused
by the distribution of $\nu_c$.

\begin{figure}
\includegraphics[width=\linewidth]{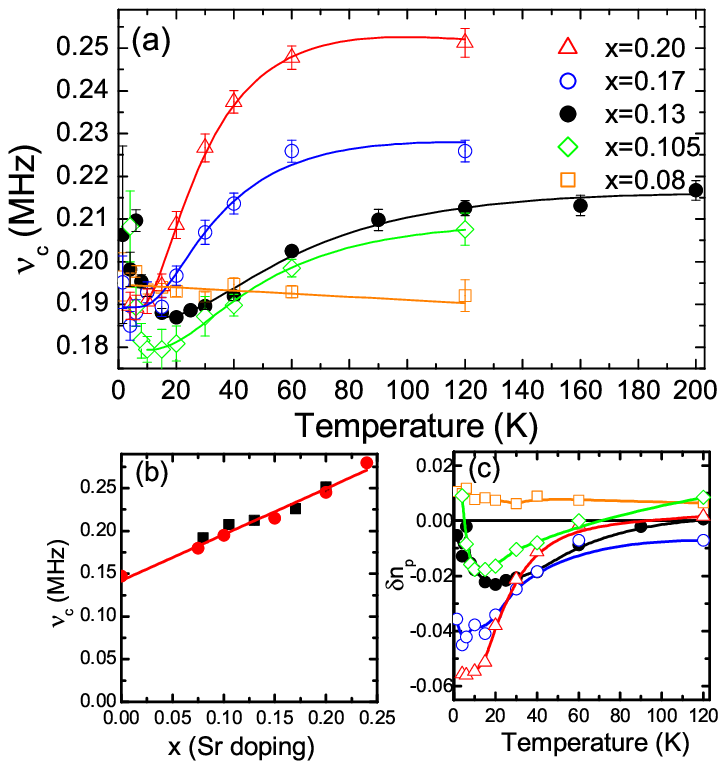}
\caption{\label{fig:nuQ}(a) The quadrupolar splitting $\nu_c$, for
several doping levels (see legend) versus temperature.  Solid
lines are guides to the eye. (b) The doping dependence of $\nu_c$
in \lescox\ (squares) and \lscox\ (circles) versus $x$ at 120K.
The solid line is given by $a + bx$, with $a$=0.142 MHz, and
$b$=0.538 MHz \cite{ohsugi}. (c) The effective change in hole
doping $\delta n_p(x,T)$, as discussed in the text. The colors and
symbols are identical to those in panel (a), and the solid lines
are guides to the eye.}
\end{figure}

The quadrupolar splitting, $\nu_c$,  is shown in Fig.
\ref{fig:nuQ}a. For $T > 120$K, $\nu_c$ varies linearly with
doping in exactly the same fashion as \lscox\ (Fig.
\ref{fig:nuQ}b). However, as seen in Fig. \ref{fig:nuQ}a, $\nu_c$
becomes strongly temperature dependent below a temperature $T_q
\sim$ 80K, in contrast to the temperature independent EFG observed
in \lscox\ or \ybcox\ \cite{zheng}.  This unexpected result is our
most important observation. One explanation for this behavior is
that the change in the EFG reflects a change in the lattice;
however there is only a $\sim0.5$\% decrease in unit cell volume
between $x$=0 and $x$=0.20, whereas the EFG increases by 70\% over
the same range \cite{takagi}. Furthermore, there is little or no
change of $\nu_c$ at $T_{\rm LT}$, suggesting that the change at
$T_q \ll T_{\rm LT}$ is unrelated to modifications of the lattice.

In fact, the dominant contribution to the EFG  at the planar O
nucleus  is the on-site charge distribution of the holes in the
oxygen 2p orbitals \cite{zheng,adrian}.  As the doping, $x$,
increases, the number of holes in the 2p orbitals, $n_p(x)$,
increases, and $\nu_c$ increases linearly with $x$: $\nu_c = a + b
x$ (Fig \ref{fig:nuQ}b). The decrease in $\nu_c$ in \lescox\
likely reflects a decrease in the hole concentration at the
in-plane oxygen sites. For concreteness, we assume that $n_p(x)=
n_p^0+ x / 2$, where $n_p^0$ is the number of holes in the
p-orbitals in the absence of Sr doping, and calculate $\delta
n_p(x,T) = \delta\nu_c(x,T)/2b = \left[\nu_c (x,T) - (a+bx)
\right] / 2b$ (Fig. \ref{fig:nuQ}c). The validity of this
assumption is supported by a recent analysis of NMR and atomic
spectroscopic results indicating that the doped holes reside
almost exclusively on the planar oxygen, and that the EFG is
linearly proportional to $n_p$ \cite{haase}. As seen in Fig.
\ref{fig:nuQ}c, $\delta n_p$ decreases by up to $\sim$ 0.05 below
$T_q$.  The increase in $\nu_c$ for $x=0.13$ is probably an
artifact due to the difficulty in fitting spectra dominated by
large magnetic broadening at low temperatures (Fig.
\ref{fig:spectra}).

\begin{figure}
  \centering
\includegraphics[width=\linewidth]{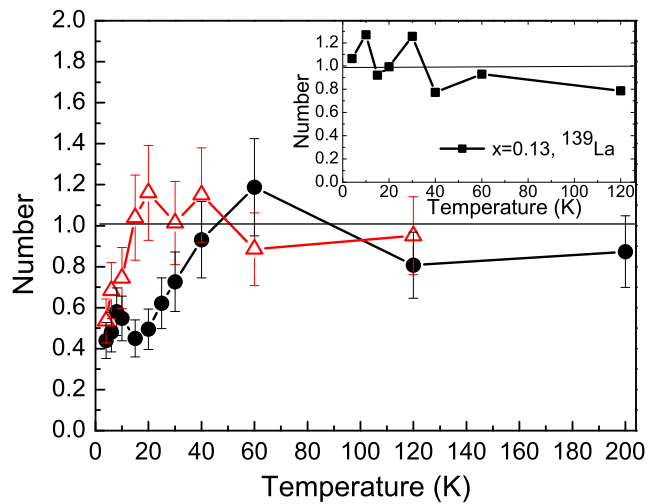}
\caption{\label{fig:wipeout}  The number of visible O nuclei
versus temperature, $N_0 \sim (I\times T) e^{2\tau/T_2}$, where
$I$ is the integrated spectral intensity, $T$ is the temperature,
$\tau$ is the pulse spacing in the NMR echo sequence, and $T_2$ is
the spin echo decay constant, for $x=0.13$ ($\bullet$) and
$x=0.20$ ($\triangle$). The data have been normalized to unity for
high temperatures.   INSET: The number of visible La nuclei versus
temperature for the $x=0.13$ sample.}
\end{figure}

\begin{figure}
  \centering
\includegraphics[width=\linewidth]{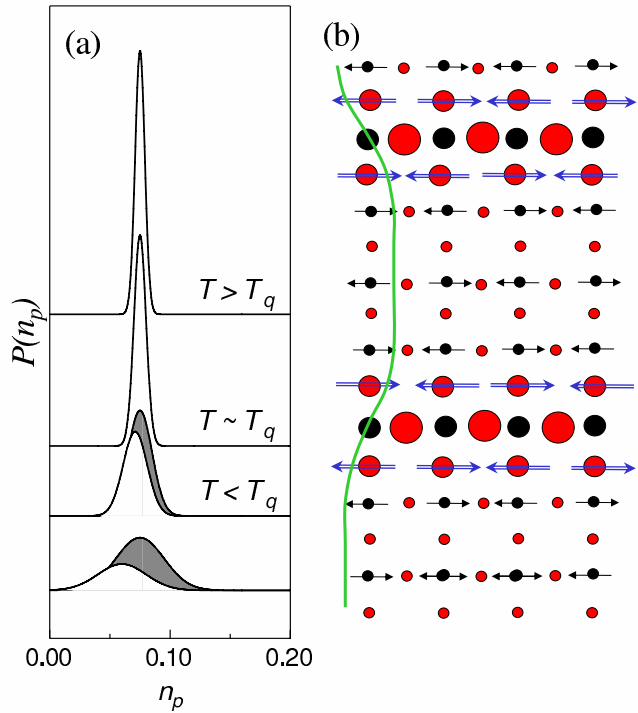}
\caption{\label{fig:picture} (a) Temperature dependence of the
planar hole doping distribution. As temperature decreases, the
width of the distribution increases monotonically, but the mean
remains temperature independent.  For $T < T_q$, a fraction of the
sites at the upper end of this distribution are wiped-out (shaded
regions), so the observed doping decreases. (b) Schematic diagram
of a stripe: the black circles and arrows represent the Cu sites
and spins, and the red circles represent the oxygens.  The
diameter of the circle is proportional to the local hole density.
The blue arrows at the oxygen sites are the local hyperfine field,
and the green line represents $\mathbf{S}(\mathbf{r})$.}
\end{figure}

The broad linewidths of the satellite peaks ($n=\pm1,\pm2$) seen
in Fig. \ref{fig:spectra} reveal a substantial distribution of
$\nu_c$, which is similar in magnitude to that observed in \lscox\
\cite{haase2} and may be attributed to a distribution of local
hole densities, $\mathcal{P}(n_p)$.  The decrease in $\delta n_p$
(Fig. \ref{fig:nuQ}b) implies either (i) the center $\langle
n_p\rangle$ of the distribution $\mathcal{P}(n_p)$ is reduced
below $T_q$, or (ii) $\langle n_p\rangle$ remains temperature
independent, but the upper end of the distribution does not
contribute to the NMR signal below $T_q$. To distinguish between
these two scenarios, we measured the absolute strength of the NMR
signal as a function of temperature
\cite{curroPRL,teitelbaumwipeout}. The temperature dependence of
the measured number of spins, $N_0$, shown in Fig.
\ref{fig:wipeout}, reveals a reduction of up to $\sim$ 50\% in the
NMR signal below $T_q$, suggesting the latter scenario. Although
precise measurements of $N_0$ are difficult due to the partial
overlap with the La spectrum, $N_0$ and $\delta n_p$ clearly
exhibit similar trends.  This interpretation has been confirmed by
recent near-edge X-ray absorption fluorescence spectroscopy
(NEXAFS) results, which indicate that $\langle n_p\rangle$ remains
temperature independent in \lescox\ \cite{fink,kroll}. We conclude
that in contrast to the behavior observed in superconducting
\lscox, the O sites located in regions of higher local hole doping
in \lescox\ do not contribute to the NMR signal, via a wipeout
mechanism discussed below. Fig. \ref{fig:picture}a illustrates
schematically how $\mathcal{P}(n_p)$ varies with temperature.

The wipeout of the O NMR signal from regions with higher hole
dopings provides concrete evidence for a spatial correlation
between the local charge modulation and the spin structure. NS
studies have shown that the fluctuating Cu spins in \lscox\ and
\lrescox\ are antiferromagnetically correlated, with a long-range
spatial modulation, $\mathbf{S}(\mathbf{r})$, that gives rise to
nodes, or domain walls, approximately every 4 lattice constants
for $x\gtrsim1/8$ \cite{tranquadaNature}. The oxygen nuclei
located adjacent to these nodes experience a large, slowly
fluctuating hyperfine field, $H_{\rm hyp}  = C
\left[\mathbf{S}(\mathbf{r} - a\hat{x}/2) - \mathbf{S}(\mathbf{r}
+ a\hat{x}/2)\right] \propto \nabla_x \mathbf{S}(\mathbf{r})/a$
(Fig. \ref{fig:picture}b). The nuclear spin lattice relaxation
rate, \slrr, of these sites is proportional to $H_{\rm hyp}^2$,
and reaches a maximum when the Cu spin fluctuation rate
($\tau_c^{-1}$) is of the order of the nuclear Larmor frequency
($\omega_L\sim 43$ MHz) \cite{CPSbook}. We estimate $T_1^{-1} \gg
(1\mu{\rm s})^{-1}$ at 10K for O sites close to a node in
$\mathbf{S}(\mathbf{r})$, whereas the time window of the NMR
spectrometer is on the order of 10$\mu$s \cite{T1calc}. Such sites
will relax too quickly to contribute to the NMR spin echo, and
will be wiped out \cite{curroPRL}. Conversely, the oxygen nuclei
located far from the domain walls where $\nabla
\mathbf{S}(\mathbf{r})$ is small experience a slower \slrr\ (the
neighboring Cu spins are locally commensurate), and thus will
contribute to the NMR spin echo signal. The onset temperatures
measured by $\delta n_p$ and $N_0$ may differ, however, for there
is not a one-to-one correspondence between the localization of a
hole along a stripe (a phenomenon driven by the local charge
distribution) and the wipeout of one oxygen site (a phenomenon
driven by the dynamics of the local spin inhomogeneity).  Although
we have no direct information about the long range topology
$n_p(\mathbf{r})$ and $\nabla \mathbf{S}(\mathbf{r})$, we can
conclude that they must be spatially correlated, otherwise there
would be no reduction of the apparent hole doping, the spectral
intensity would be reduced for all local hole dopings, and $\nu_c$
would be temperature independent, in contrast to our observations.

Two- and three-band Hubbard model calculations can shed light on
the spatial dependence of the hole doping and the spin structure
in stripe-ordered systems \cite{zaanen,martin}. These calculations
show increased hole density for the O sites in and adjacent to the
charge stripes, whereas in the intervening antiferromagnetic
regions the hole density on the O sites is equal or close to that
in the undoped material (see Fig. \ref{fig:picture}b), in
agreement with our observations of $\nu_c$. Furthermore,
$\mathbf{S}(\mathbf{r})$ vanishes at the charge stripe, and
undergoes a phase change of 180$^{\circ}$. Therefore, $\nabla
\mathbf{S}(\mathbf{r})$ is largest in the vicinity of the domain
wall, and the oxygen sites adjacent to these nodes are exactly
those with the highest hole density and are wiped out. The
observation that the hole density is greatest in regions where
$\nabla \mathbf{S}(\mathbf{r})$ is largest strongly supports the
idea of charged domain walls first discussed by Zaanen
\cite{zaanen}.

Since NMR is a local probe, we cannot determine whether
$n_p(\mathbf{r})$ is randomly distributed in amorphous islands, as
observed in recent STM experiments on
Bi$_2$Sr$_2$CaCu$_2$O$_{8+\delta}$ \cite{lang}, or exhibits the
long-range correlations characteristic of a stripe lattice.
However, we can make some general observations about the topology.
If we assume that the holes segregate into islands with a
characteristic radius $r_0$, and that the wipe-out occurs for
sites on the boundaries of the islands where $\nabla
\mathbf{S}(\mathbf{r})$ is largest, then in order to account for a
loss of $\sim$ 50\% of the sites we estimate that $r_0 \sim
15$\AA, a value of the same order as measured by STM \cite{lang}.
Note, however, that $T_c \lesssim 10$K, whereas $T_q \sim 80$K, so
a scenario of localized superconducting islands is inconsistent
with our results. Secondly, we note that approximately 25\% of the
O sites are wiped out for site-centered charge stripes (Fig.
\ref{fig:picture}b), whereas close to 50\% are wiped out for
bond-centered stripes \cite{scalapino}, or for a 2D checkerboard
pattern \cite{hoffman}. However, this number is strongly dependent
on the details of the long range topology and the width of the
domain walls, and the low precision inherent in the nature of
these measurements precludes any definite conclusions about these
scenarios.

In summary, we have observed a decrease in the EFG at the O sites
in \lescox\ that can be understood in terms of a spatial
correlation between the local hole doping and the domain walls of
the spin modulation.  Both the LTO (\lscox) and the LTT (\lescox)
phases exhibit similar hole distributions \cite{singer}, but the
glassy spin freezing present in the latter gives rise to the
wipeout of O sites near the domain walls. The fact that the widths
of the hole distributions are similar in both systems is
surprising, since it implies that some form of charge
inhomogeneity is present in both the LTO and LTT phases, at least
on the time scale of the NMR experiments, and contradicts the idea
that the LTT phase pins the spatially fluctuating charge stripes
\cite{tranquadaglass}. Rather, the LTT phase suppresses the spin
fluctuations. We speculate that charge inhomogeneity is present in
both systems \cite{bishop}, but that the spin dynamics are
strongly affected by the LTT phase.

We thank A. Bishop, J. Haase, and J. Tranquada for enlightening
discussions. This work was performed at Los Alamos National
Laboratory under the auspices of the U.S. Department of Energy.
The work at Brookhaven was supported by the Office of Science,
U.S. Department of Energy under Contract No. DE-AC02-98CH10886.


\begin{thebibliography}{34}
\bibitem[{api()}]{apicalnote}
\bibinfo{note}{The apical O signal is suppressed since \slrr\ is much longer
  than the experimental repetition time.}
\bibitem[{T1c()}]{T1calc}
\bibinfo{note}{\slrr\ reaches a maximum value of $\gamma^2 H_{\rm
  hyp}^2/\omega_L$ when $\omega_L\tau_c\sim 1$, for temperatures on the order of
  10K \cite{curroPRL}. Assuming $H_{\rm hyp}=C\mu_{\rm Cu}$ at a domain wall
  (see Fig. \ref{fig:picture}) and $\mu_{\rm Cu}\approx 0.5\mu_B$ we have
  $T_{1,\rm{max}}^{-1} \sim (10^{-9} {\rm sec})^{-1}$.}
\end{thebibliography}

\end{document}